\documentclass[pdftex,dvipsnames,usenames,aps,superscriptaddress,amsmath,showkeys,twocolumn,prd,10pt]{revtex4-2}

\usepackage{lineno,hyperref,color}

\usepackage{graphicx}
\usepackage{amssymb}
\usepackage{url}
\urldef\Weather20A7\url{http://geoservices.knmi.nl/viewer2.0/?srs=EPSG%3A3857&bbox=-725.7203842048766,6293973.625155548,1220725.7203842048,7406026.374844452&service=http%3A%2F%2Fgeoservices.knmi.nl%2Fcgi-bin%2FRADNL_OPER_R___25PCPRR_L3.cgi%3F&layer=RADNL_OPER_R___25PCPRR_L3_COLOR%24image%2Fpng%24true%24rainbow%2Fnearest%240.83%240&selected=0&dims=time$2020-08-14T14:15:00Z&baselayers=streetmap}
\def\figref#1{Fig.~\ref{fig:#1}}
\def\figlab#1{\label{fig:#1}}  
\def\tabref#1{Table~\ref{tab:#1}}
\def\tablab#1{\label{tab:#1}}  
\def\eqref#1{Eq.~(\ref{eq:#1})}

\def\eqlab#1{\label{eq:#1}}
\newcommand*{\secref}[1]{Section~\ref{sec:#1}}
\newcommand*{\seclab}[1]{\label{sec:#1}}

\newcommand{\Omit}[1]{}

\RequirePackage[normalem]{ulem}

\def\KVI{University of Groningen, KVI Center for Advanced Radiation Technology, Groningen, The Netherlands}
\def\Kapt{University Groningen, Kapteyn Astronomical Institute, Landleven 12, 9747 AD Groningen, The Netherlands}
\def\AIVUB{Astrophysical Institute, Vrije Universiteit Brussel, Pleinlaan 2, 1050 Brussels, Belgium}
\def\IIHE{Interuniversity Institute for High-Energy, Vrije Universiteit Brussel, Pleinlaan 2, 1050 Brussels, Belgium}

\def\IMAPP{Department of Astrophysics/IMAPP, Radboud University Nijmegen, Nijmegen, The Netherlands}

\def\ASTRON{Netherlands Institute for Radio Astronomy (ASTRON), Dwingeloo, The Netherlands}

\def\UNH{Department of Physics and Space Science Center (EOS), University of New Hampshire, Durham NH 03824 USA}

\def\Erl{Erlangen Center for Astroparticle Physics, Friedrich-Alexander-Univerist\"{a}t Erlangen-N\"{u}rnberg, Germany}
\def\KIT{Institut for Astroparticle Physics, Karlsruhe Institute of Technology(KIT), P.O. Box 3640, 76021, Karlsruhe, Germany}
\def\DESY{DESY, Platanenallee 6, 15738 Zeuthen, Germany}

\begin{document}

\title{Time resolved 3Dinterferometric imaging of a section of a negative leader with LOFAR}

\author{O.~Scholten}
    \email[Correspondence email address: ]{O.Scholten@rug.nl}
    \affiliation{\Kapt,\KVI,\IIHE}

\author{B.~M.~Hare}
    \affiliation{\Kapt}  
\author{J.~Dwyer} %
   \affiliation{\UNH} 
\author{N.~Liu} %
   \affiliation{\UNH}   
\author{C.~Sterpka} %
   \affiliation{\UNH}  
\author{S.~Buitink}    \affiliation{\IMAPP} \affiliation{\AIVUB}  
\author{T.~Huege}     \affiliation{\KIT} \affiliation{\AIVUB} 
\author{A.~Nelles}    \affiliation{\Erl} \affiliation{\DESY}  
\author{S.~ter Veen}    \affiliation{\ASTRON}   

\date{\today}

\begin{abstract}
We have developed a three dimensional (3D) interferometric beamforming technique for imaging lightning flashes   using Very-High Frequency (VHF) radio data recorded from several hundreds antennas with baselines up to 100~km as offered by the Low Frequency Array (LOFAR). The long baselines allow us to distinguish fine structures on the scale of meters while the large number of antennas allow us to observe processes that radiate at the same intensity as the background when using a time resolution that is close to the impulse-response time of the system, 100~ns.
The new beamforming imaging technique is complementary to our existing impulsive imaging technique.
We apply this new tool to the imaging of a four stepped negative leaders in two flashes. For one flash, we observe the dynamics of coronal flashes that are emitted in the stepping process. Additionally, we show that the intensity emitted in VHF during the stepping process follows a power-law over 4 orders of magnitude in intensity for four leaders in two different lightning storms.
\end{abstract}


\maketitle

\section{Introduction}


Due to the fact that it is unpredictable, violent, and often hidden in clouds, the common phenomenon of lightning still harbors many secrets such as the conditions for initiation and the processes that allow the formation of conducting ionized channels called leaders, just to name a few. During any part of its development a lightning flash emits copious amounts of electromagnetic radiation over a very broad frequency spectrum, ranging from static electric fields to gamma-ray emission. For this work we use the emission in the Very-High Frequency (VHF) band to image the three dimensional (3D) lightning development in time. The major advantage of VHF-imaging over, for example, optical imaging, is that VHF is not obscured by the presence of clouds allowing for large distance observations.

The imaging of lightning in three dimensions using VHF-interferometric-based methods has a history dating back to the early early work of \citep{Mardiana:2002} where 3D is reached by combining angular measurements from two different stations. This work was followed up by others, such as \cite{Akita:2014,Stock:2014,LiuHY:2018} and culminated recently in the work of \cite{Jensen:2021} reaching a resolution of 200~m using 6 antennas.

We have developed similar techniques for imaging lightning with LOFAR \cite{Haarlem:2013}, which has resulted in much improved resolution. In this previous work we had used the signals of several hundred antennas for 3D imaging \cite{Hare:2019}. In this impulsive-imaging approach the time of the peak in the cross correlation between a pulse on the reference station and pulses in other stations is used in a chi-square fitting procedure to find the most likely location for the source. This impulsive imager was optimized a step further in \cite{Scholten:2021-init} and can find over 200 source locations per millisecond where meter-scale accuracy is reached.

Despite the history in the lightning field of naming these previous techniques as "interferometry", extremely few of them \cite{Stock:2014-PhD, Tilles:2019} use actually true interferometry in the sense of coherently summing the electric fields measured by many antennas.
In this work we show the results of using a true interferometric beamforming technique for imaging lightning in 3D using hundreds of LOFAR antennas with maximal baselines of 100~km.
In beamforming the signals from hundreds antennas are coherently added and the imaged volume is rastered into voxels, volumetric pixels.
Where most interferometric approaches integrate over considerable time periods, we divide the beamformed time trace for each voxel in narrow slices. A slice may be as small as 100~ns, the impulse-response time of the LOFAR antennas. For each time slice and each voxel the coherent intensity is calculated, thus forming a 4D (three space and one time dimension) intensity image. For each time slice the position of maximal coherent intensity is taken as the position of the source. In spite of the high time resolution we still resolve sources that emit with intensity well below the noise limit, where the noise level of a single antenna is dominated by the galactic background \cite{Mulrey:2019}. Since we can reach a time resolution of the order of the impulse-response time of our system we name this procedure Time Resolved Interferometric 3-Dimensional (TRID) imaging. In \secref{Method} the TRID imaging technique is discussed in detail.

Lightning, being an electric discharge phenomenon, has positively and negatively charged propagating structures, called leaders. While the positive leaders propagate in a rather continuous way, the negative leaders, in contrast, propagate discontinuously through a stepping process. In this stepping process copious amounts of VHF are emitted making them ideal test cases for our imaging method. As a first application we thus have used TRID imaging to investigate negative stepped leaders in \secref{NegLead}. The main draw-back of interferometric 3D imaging is that it is compute-intensive, thus without large dedicated compute resources only a relatively small region of space can be imaged. In \secref{LargeScale} we explain our two-step approach where the entire lightning flash is imaged with the impulsive imager \cite{Scholten:2021-init} and from this image the interesting parts are selected that are subsequently imaged using the TRID imager.
We show in \secref{Dynamics} that the TRID imager allows to resolve the dynamics of a corona flash for some negative leaders. As a second application we show in \secref{PowerLaw} that the intensity spectrum of VHF-pulses emitted from negative leaders follows an almost perfect inverse-square power-law over four orders of magnitude (with a scaling exponent of $-1.8$).
We end with a short discussion in \secref{summ}.

\section{Methods}\seclab{Method}

The LOFAR infrastructure consists of several thousand antennas distributed in stations over much of Europe which a dense core in the Netherlands. For the present analysis we use the antennas installed in the Netherlands. The LOFAR antennas are arranged in stations each containing 96 Low Band Antennas (LBA, inverted v-shaped dipoles operating in the 30 -- 80~MHz band) and a similar number of High Band Antennas (100-200~MHz, not used here). For practical reasons based in the observation mode the data of 12 dipoles each from all 37 Dutch stations is used where the geographic spread is shown in \figref{LOFAR-NL} covering a logarithmic baseline distribution up to 100~km.  At the core of LOFAR, called the Superterp, the antenna stations are more closely packed. Upon a trigger the RAM buffers containing raw data from every antenna are frozen and read out over glass fiber to a central storage. This allows for off-line processing of the time traces (sampled at 200~MHz) of all antennas. The atomic clocks in the stations allow for a timing stability of a fraction of a nanosecond over the recording time span of maximally 5 seconds.

\begin{figure}[th]
\centering{\includegraphics[bb=3.5cm 3.5cm 25.5cm 25.cm,clip, width=.47\textwidth]{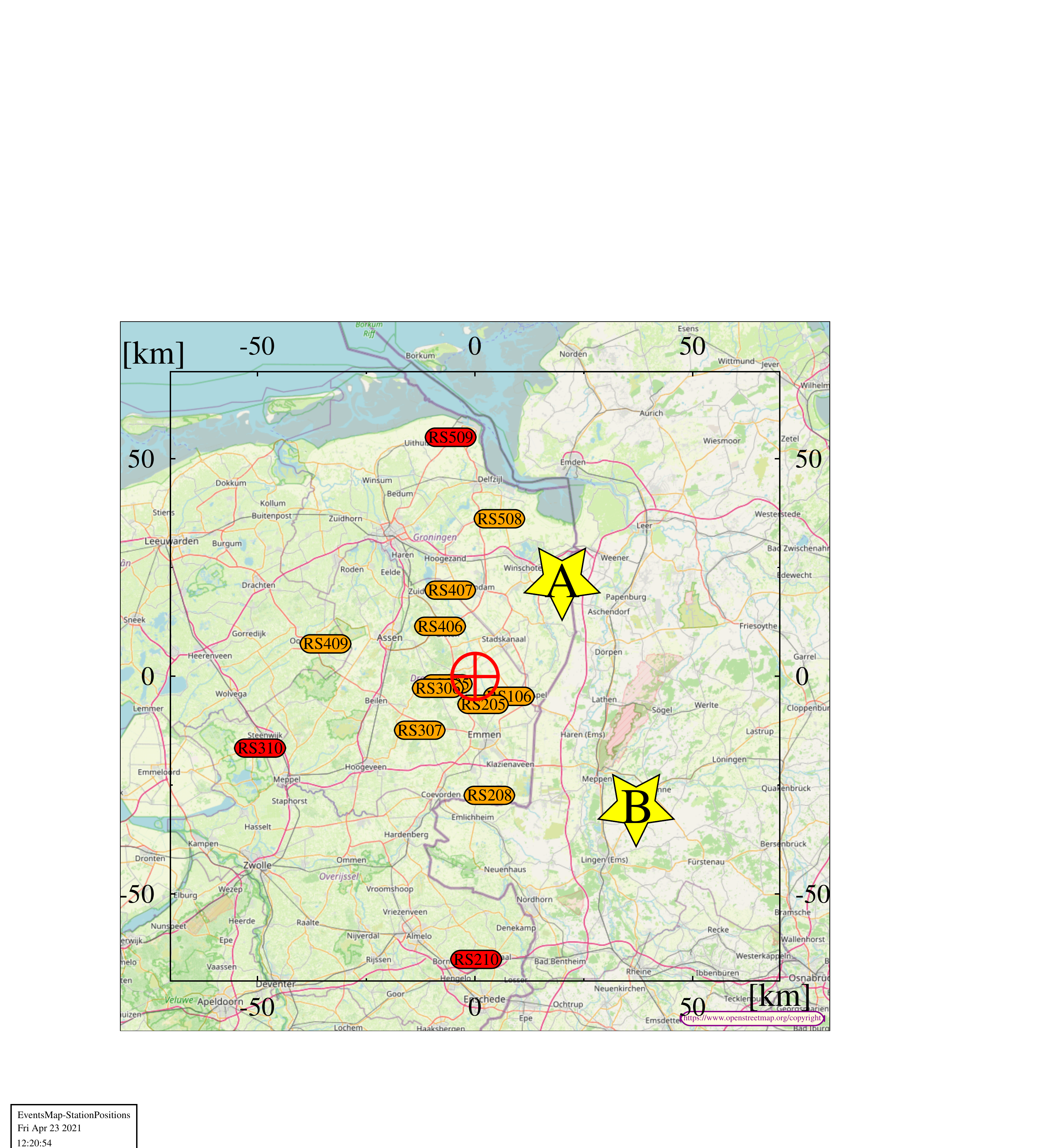} }
	\caption{Layout of the Dutch LOFAR stations. The core of LOFAR is indicated by the red $\bigoplus$\ sign while the yellow stars show the general location of flashes A and B that are  discussed in this work. The black frame indicates the general area ($140\times 140$~km$^2$) where flashes can be mapped accurately and is centered at the LOFAR core.  Stations marked in red are not included in TRID imaging (see text).}	 \figlab{LOFAR-NL}
\end{figure}

The basic implementation of beamforming imaging is relatively straightforward. The part of the atmosphere where the lightning is to be imaged is rastered into voxels, see \secref{Vox}. For each voxel the time traces of all antennas are summed while accounting for the difference in travel time to each antenna. This yields for each voxel a time trace which is the coherent sum of all antennas, as discussed in \secref{Vox}. Adding the traces for the different voxels is always performed for a fixed time span in one particular `reference' antenna, usually an antenna from the core of LOFAR. Thus we can find the source location of a particular structure seen in the trace of the reference antenna. The choice of reference antenna is not important as long as it is in the dense core.

The time trace of each voxel is cut in slices of a fixed duration. The slicing is done such that they are synchronous when correcting for the signal travel time from the reference antenna to the voxel. For all slices the coherent intensity is determined resulting in a voxelated intensity profile for each time slice, similar to what is shown in \figref{B-NLa-166}. For each time slice the maximum of this voxelated intensity profile is determined and used as the source location, as discussed in more detail in \secref{Max}.
Plotting the position of all sources will result in the beam-formed (or interferometric) image of a segment of the flash in space and time where the length of the time slices determines our final time resolution since any detail within a slice is summed over. In the following the different steps involved in this procedure are discussed in more detail.


\subsection{Space \& time grids and summing time traces}\seclab{Vox}

The part of the atmosphere that is of interest is rastered into voxels. We have found that this is most conveniently done in spherical coordinates with the center at the reference antenna in the dense LOFAR-core. In this way we can more easily account for the fact that our resolution in the two transverse directions is much better than in the radial direction. The optimal grid-spacing is dependent on the areal spread of the antenna locations. For the example discussed in this work we include antennas within 50~km distance from the core where the antennas are distributed on an irregular logarithmic grid with a dense core that is designed to be optimal for astrophysical applications, see \cite{Haarlem:2013} and \figref{LOFAR-NL}. The voxel grid is chosen such that the maximal time shift for any antenna for two bordering voxels is about 1~ns, our time calibration accuracy. This implies for the examples discussed in  \secref{NegLead} a grid spacing of 0.003$^\circ$ in the azimuth angle ($\hat\phi$), 0.01$^\circ$ in elevation angle ($\hat\theta_e$, upwards from the horizon), and 10~m in the radial direction ($\hat{R}$). For a source at 50~km distance this implies a grid spacing of about 1~m transverse to the line of sight (from the reference antenna) and 10~m along the line of sight. A typical grid used in  \secref{NegLead} spans over $(60 \times 30 \times 20)$ voxels in $(\hat\phi,\,\hat\theta_e,\,\hat{R})$.


A section of the time trace in the reference antenna is selected with typically a length of 0.3~ms. For each voxel the
time traces of all antennas are added after time shifting them to account for the travel time differences of a signal from the voxel. This yields the coherent (or beam formed) time trace for that voxel. All coherent voxel time traces are evaluated at identical times for the reference antenna. 

The actual time shift of the time traces is done in Fourier space by applying a frequency-dependent phase shift to the trace for each antenna. After summation, the traces are transformed back to the time domain. The actual antenna time traces are taken longer than the required 0.3~ms to be able to account for the maximal relative time shift over the full voxelated volume.
We thus obtain for each voxel a coherent time trace, with all antennas contributing, of 0.3~ms length with additional trailing ends where there will be partial coherence. These trailing end are excluded from the analysis.

\begin{figure}[th]
\centering{\includegraphics[ width=0.49\textwidth]{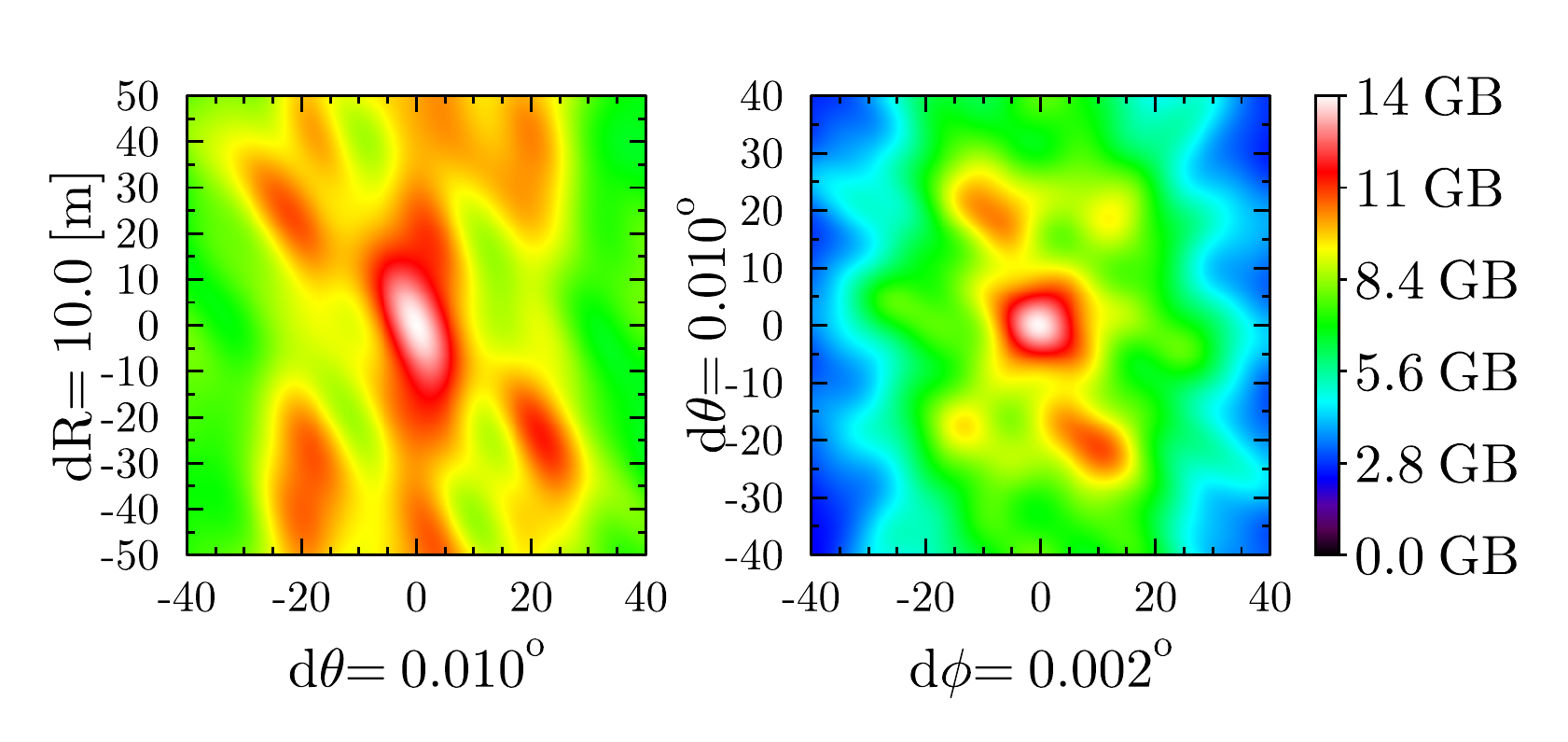} }
	\caption{A typical interferometer image for Flash B. The axes show the grid in voxel number where $(0,0,0)$ is the center voxel in the image cube.  Both images are made for a thin (one voxel thick) section through the center voxel. To show the general features of the intensity plot, the hyper cube is chosen such that the maximum intensity is at the center voxel, which generally is not the case. The typical length scale for $d\phi$ and $d\theta_e$ is about 1~m. The intensity unit [gb] is explained in \secref{Cal}. }	 \figlab{B-NLa-166}
\end{figure}

The intensity may be integrated over the full time trace of 0.3~ms which washes out all time dependence over the integration range.
However, a much better time resolution is generally required in order to image the dynamics of the flash.
To obtain this, we cut the time trace from each voxel into narrow slices and we integrate the power over each slice for every voxel to obtain the voxelated intensity profile for each slice, giving an image like \figref{B-NLa-166}. The optimal slice length depends on the research question of interest, taking it short resolves  finer time structures but makes the imaging more noisy, taking it long  improves the signal-to-noise ratio, and thus the sensitivity, but increases the chance that multiple sources appear simultaneously and are confused.
The smallest time step where subsequent time-frames are considered to be independent is the width of the impulse response function.
The impulse-response time is determined by the frequency dependence of the LOFAR antennas, amplifiers, and filters. An easy estimate of this is obtained from the full width at half maximum of the narrow peaks in the spectrum, about 50~ns.
As shown in \secref{NegLead} very stable results are obtained even for a time slice as short as 100~ns, which is a little longer than the impulse response time.
Since we have shown that we can reach down to close to the impulse-response time of the system we name this Time-Resolved Interferometric 3D (TRID) imaging.

\subsection{Antenna weighting}\seclab{AntWeight}

Since the array of antennas has an extent that is comparable to the distance to the source region, the received signal strength varies greatly over the array. If this is not taken into account in the analysis one would effectively use the antennas that are near the source region only and thus potentially use only a fraction of the imaging accuracy of the array as a whole. To improve this aspect of the performance we include weighting factors for the antennas to compensate for the imbalance in signal strength. In interferometry applications for astronomy (2D, not 3D imaging) there is a considerable amount of literature regarding the advantages and disadvantages of different weighting schemes, see \cite{Briggs:1999, Yatawatta:2014} for comprehensive reviews.

For this work we have chosen for a weighting scheme that compensates in lowest order the signal strength variations in the antennas due to their distance from the source, where the weight is capped for antennas far from the source for which the signal to noise ratio becomes unfavorable. In addition it takes into account the fact that the largest density of antennas is at the core.

The amplitude of an emitted signal drops inversely proportional to distance $R_{as}=\sqrt{D_{as}^2+h_s^2}$, where $D_{as}$ is the distance from a certain antenna to the source in the horizontal plane and $h_s$ denotes the altitude of the source. For this weighting scheme  it is assumed that the antennas are all in a horizontal plane since that part of the Netherlands is flat to a good approximation, in all other interferometry calculations the proper 3-dimensional locations of the antennas are used. Another important factor to account for is the antenna gain that depends on the azimuth $\phi$ and elevation $\theta_e$ angles of the source with respect to the antenna. The antenna gain vanishes for sources at the horizon and thus (since it is an analytic function of the angles) depends on $\theta_e$ as $\sin{(\theta_e)}$. For almost all cases of interest the sources are at small elevation angles for which we can assume $\sin{(\theta_e)}\approx h_s/R_{as}$ where the antenna position enters in $R_{as}$ only.

The antenna gain depends also on $\phi$ as well as on the polarization of the signal. Since an analysis of the polarization of the radiation falls outside the scope of this work as it would make the analysis considerably more complicated, we have ignored this dependence. Even though the antennas are not in a small azimuthal angle range from the source (full opening angle about 60$^\circ$, see \figref{LOFAR-NL}) we still achieve high-quality results.

Combining the $1/R_{as}$ drop of signal strength with the proportionality of the antenna gain to $\sin{(\theta_e)}$, we obtain an antenna weighting factor
\begin{equation}
W_a=\left\{
  \begin{array}{lr}
    R_{as}^2/R_{rs}^2 & \rm{if} \quad R_{as}^2/R_{rs}^2 < W_{\rm max}\\
     W_{\rm max} & \rm{if} \quad  R_{as}^2/R_{rs}^2 \ge  W_{\rm max}
  \end{array} \;. \eqlab{W}
\right.
\end{equation}
where the weight is normalized to unity for the reference antenna at a distance of $R_{rs}$. In addition the weight is capped to a maximum of  $W_{\rm max}=1.2$ for distant stations where the signal to noise ratio is getting worse. It is conceivable that a different choice for $ W_{\rm max}$, for example by making it dependent of the signal-to-noise ratio, will improve the image accuracy.

\subsection{Antenna calibration and Intensity}\seclab{Cal}

The antenna timings are calibrated for each flash following the procedure outlined in \cite{Hare:2019,Scholten:2021-init}. Per flash 20  -- 30 bright stand-alone pulses are selected from the whole flash where care is taken that their source locations roughly cover the extent of the flash. For all sources the location as well as the antenna timings are searched in a simultaneous fit using the search algorithm discussed in \cite{Scholten:2021-init}.

Since for interferometry the stability of the phase of the signal over all antennas is important, we perform an additional check of this phase for a single distinct pulse in the spectrum selected for TRID imaging. If this phase is off by more than 90$^\circ$ the antenna will not be used. Usually this eliminates less than 1\% of all antennas. This may also result in eliminating a whole station from the analysis for a particular flash. The latter could be due to the fact that we have ignored the azimuthal angle dependence of the antenna function.

The gain of each antenna is calibrated by normalizing the noise level to unity. The noise level is determined from the parts of the recorded trace for which there is no lightning activity detected within 0.3~ms.  Since this noise level is largely due to non-terrestrial sources, i.e.\ radiation produced by the Milky Way, we name this the Galactic Background [gb] level.

The coherent or interferometric intensity is calculated as the intensity of the signal from the source as received by the reference antenna and is expressed in units of [gb]). It is tacitly assumed that the azimuth-angle dependence of the antenna gains is limited and is effectively averaged-out. The elevation-angle dependence is to a large extent accounted for by the weighting factors (see \secref{AntWeight}) when calculating the coherent intensity.

All LOFAR antennas used in this work are inverted v-shape dipoles where X- (Y-) dipoles, corresponding to odd (even) numbered antennas, are oriented in the NE-SW (NW-SE) plane. Our analysis is performed separately for X- and Y-dipoles since they differ significantly in their sensitivity to polarized radiation. The antenna function (the Jones matrix, specifying for each dipole and polarization the gain depending on angle and frequency) has to be used \cite{Haarlem:2013} to convert the measured intensity to the absolute intensity of the source, which we have not done in this work. The absolute calibration of the LOFAR antennas is performed in \cite{Mulrey:2019}.
Extracting intensities (and even polarization) of a source is considerably simplified if it is assumed that the source is point-like emitting a delta-pulse (flat frequency spectrum over our frequency range) and linearly polarized, however such an analysis falls outside the scope of the present work.


\subsection{Determining the location of maximal intensity}\seclab{Max}

\figref{B-NLa-166} shows a typical intensity profile for a time slice of 10~$\mu$s. For this particular one the coordinates of the hypercube have been chosen such that the peak intensity is at the center. Generally this is of course not the case. The procedure used to determine the position of the maximum is discussed in this section. Besides the main maximum there are several secondary maxima, due to so-called side-beams, a well known feature in interferometric imaging. The positions and intensities of these side beams is determined by the antenna configuration. The LOFAR irregular lattice of antennas suppresses the intensities of these side beams to a large extent. Secondary maxima could also be due to background noise or due to multiple sources in the same time slice.
It is straightforward to determine the voxel with maximal intensity. To reach an inter-voxel accuracy we have implemented two interpolation procedures, quadratic and barycentric interpolation.
\begin{description}
\item[Quadratic] The intensity of the voxels around the voxel with the maximum intensity are fit with a paraboloid,
    \begin{equation}
    I_p(\vec{x})=I_0 + \sum_i\left[\frac{1}{2} A_i x_i^2 + B_i x_i\right] + \sum_{i<j} R_{i,j} x_i x_j \;,
    \end{equation}
    where $x_i$ is the $i^{th}$ grid coordinate, $i=1,2,3$. The coefficients $I_0$, $A_i$, and $R_{i,j}$ are fitted to the grid points bordering the maximum. The inter-voxel maximum $\vec{x_p}$ is taken at the point where the paraboloid reaches its maximum.
\item[Barycentric] The barycentric maximum is calculated as
    \begin{equation}
    \vec{x}_b= \sum_{\vec{x}} \left[(I(\vec{x})-I_{\rm Th})\, \vec{x} \right] /  \sum_{\vec{x}} \left[(I(\vec{x})-I_{\rm Th}) \right] \;,
    \end{equation}
    where the sum runs over all voxels with an intensity exceeding the threshold value $I_{\rm Th}$. This threshold is taken as $I_0/1.2$ (where $I_0$ is the maximum voxel intensity) or the largest value of a voxel on the outer surface of the grid, whichever is larger.
\end{description}
When there are many active sources in small area, the barycentric interpolation will yield some weighted average position, while the quadratic interpolation yields the position of the strongest source.
It is observed that the details of intensity surface are complicated with small ripples (order 10\% in intensity over distances of 20~m depending on the source location) likely due to side beams. These ripples make the quadratic interpolation unstable for a coarse grid. Using barycentric interpolation these ripples are efficiently averaged yielding a properly interpolated maximum. The examples shown in \secref{NegLead} use a fine grid and for this reason the quadratic interpolation is used.


\section{Negative Leaders}\seclab{NegLead}

As an example of the possibilities offered by the new TRID imager we apply it to some normal negative leaders. These are selected as their structure is assumed to be relatively simple and one thus obtains a good insight in the intrinsic accuracy of the TRID imager. We use two negative leaders from each of two flashes, Flash A occurring on April 24, 2019 at 21:30:56.221 UTC and Flash B on August 14, 2020 at 14:14:58.669 UTC. These times are taken as $t=0$ for each flash. We show a zoomed-in view on the fine structure of one negative leader of Flash B and statistics for all four negative leaders.

The location of the two flashes with respect to the LOFAR core is shown in \figref{LOFAR-NL}. Both flashes have been imaged using the impulsive imager, where Flash A was reported on in \cite{Scholten:2021-init,Scholten:2021-RNL} and Flash B in \cite{Scholten:2021-HANL}. For the impulsive imager all stations are used, including those marked in red in \figref{LOFAR-NL}, while for TRID imaging we have excluded these stations because they are at a distance larger than 50~km from the LOFAR core. This limit on distance is taken since for further distances the signal becomes weak, or the spread in angles to the source too large. For Flash B also station RS208, RS508, and RS307 were excluded since for these the phase-stability for selected pulses could not be shown, see \secref{Cal}. 

The sections of the negative leaders which we will use for the discussion in this work are given in \tabref{NegLead}.
\begin{table}[ht]
\centering
\begin{tabular}{||c|c||r@{ -- }l|r@{ -- }l|r@{ -- }l|r@{ -- }l||}
\hline
Flash & Leader & \multicolumn{2}{c|}{time [ms]} & \multicolumn{2}{c|}{North [km]} & \multicolumn{2}{c|}{East [km]} & \multicolumn{2}{c||}{height [km]} \\
\hline \hline
A & NL-A1 & 170& 178 &    21 & 23  &  18&20 &   2.5&3.5 \\
A & NL-A2 & 177&182  &    21&23    &  21&22 &   3. & 3.5  \\
B & NL-B1 & 238& 245 & $-25$&$-25$ &  37&38 &  5.3&5.8 \\
B & NL-B2 & 540& 552 & $-34$&$-34$ &  36&37 &  4.9& 5.5 \\
\hline
\end{tabular}
\caption{The negative leader sections that are used for further analysis. \tablab{NegLead}}
\end{table}
In \secref{LargeScale} we show that while the image quality on a large scale may be somewhat better for the TRID imager as compared to the impulsive imager, this does not outweigh the additional effort in CPU power. As is shown in \secref{Dynamics} the true strength of the TRID imager lies in resolving fainter sources keeping good time resolution. Another strong point of the TRID imager is that it is complete in the sense that it will locate all sources with a strength exceeding background in the image-cube with only minor caveats. The impulsive imager is in this respect much less complete since the located sources  are not restricted by the small hypercube (which can also be an advantage) and furthermore need to be well separated from earlier and later sources. The completeness of the TRID-imaged sources allows to analyze the VHF-intensity spectrum, as discussed in \secref{PowerLaw}.

\subsection{Larger scale structures}\seclab{LargeScale}

\begin{figure*}[th]
\centering{\includegraphics[ bb=1.0cm 2.4cm 24.5cm 25.7cm,clip, width=0.49\textwidth]{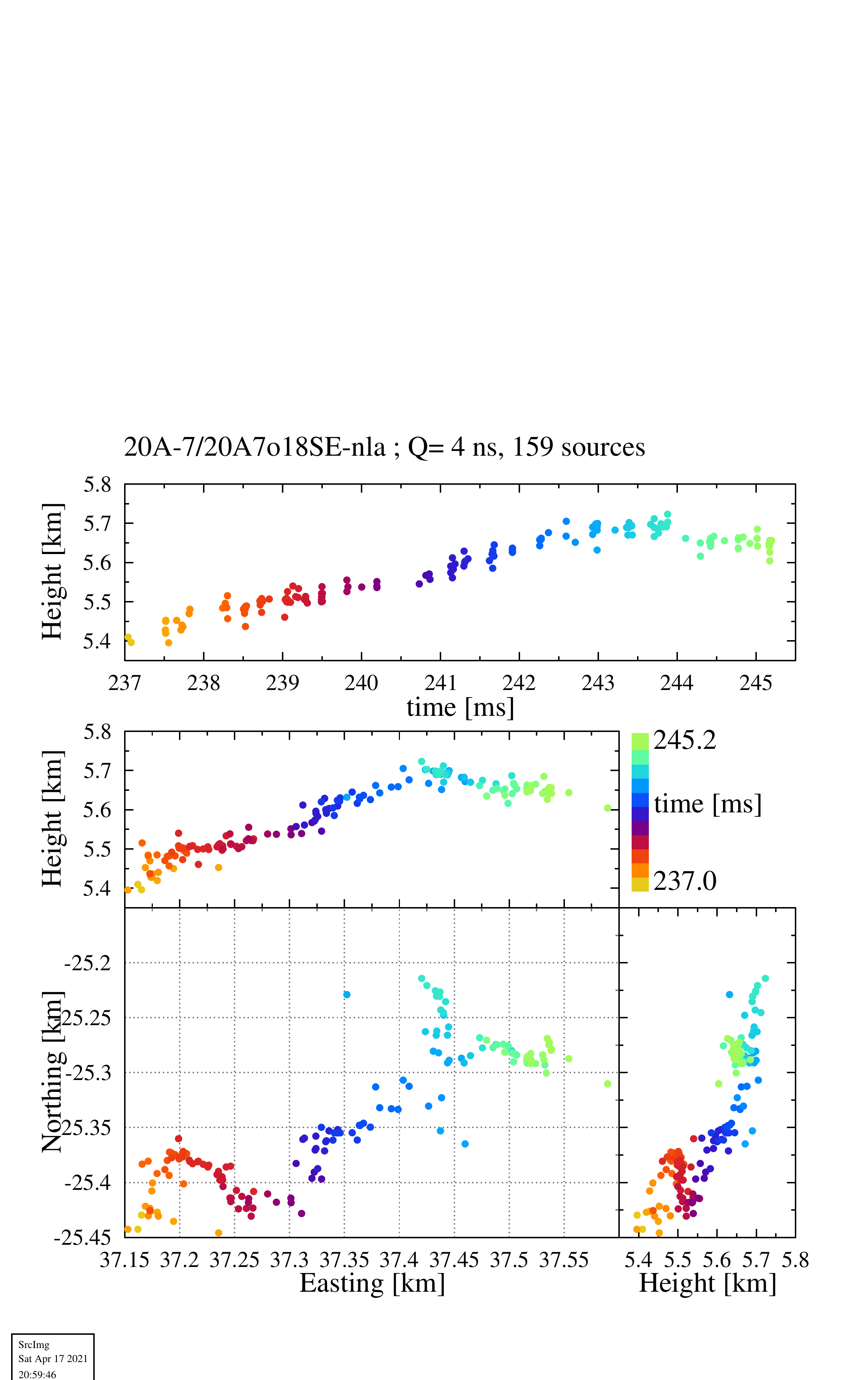} } 
\centering{\includegraphics[ bb=1.0cm 2.4cm 24.5cm 25.7cm,clip, width=0.49\textwidth]{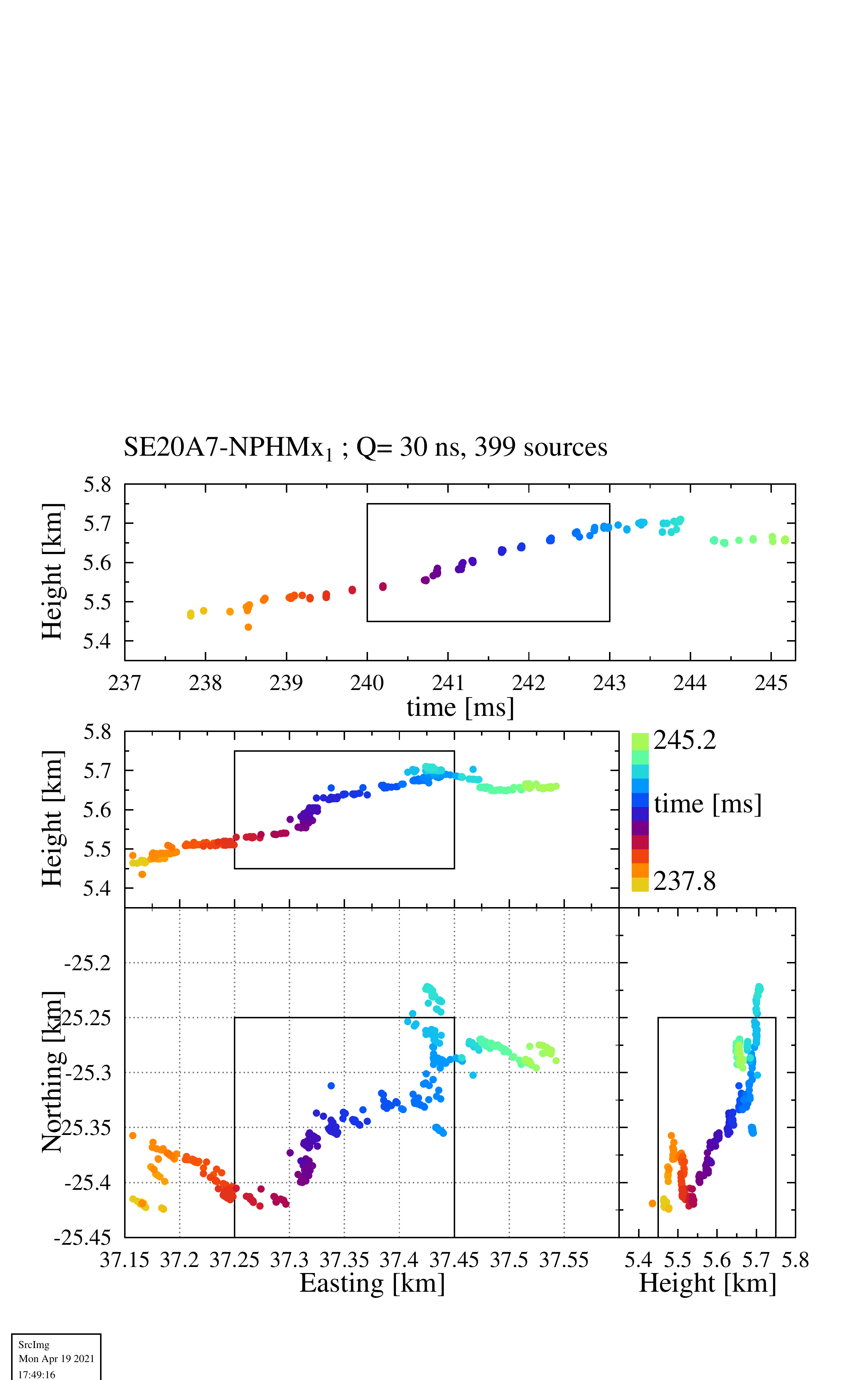}} 
	\caption{A comparison of the images for section `NL-B1' from a negative leader of Flash B (see \tabref{NegLead}) as obtained by the impulsive imager (left panels) and the TRID imager (right panels) using the X-dipoles from LOFAR. The boxes on the right indicate the zoom-in volume for \figref{Intf-NLaZM}, left. }	 \figlab{Intf-NLaH}
\end{figure*}

Interferometric imaging is rather expensive. For the present implementation it typically takes half a day on a single CPU to produce an TRID image of a volume of about 1~km$^3$ for a trace of 0.3~ms while it takes about 1 hour to image a section of 100~ms of a flash with the impulsive imager, however, since the interferometric imaging code can easily be paralleled, this problem may not be as constraining at it appears. Both estimates are for a typical LOFAR configuration with a few hundred antennas.   In \figref{Intf-NLaH} the results of the two imaging procedures are compared for a section of about 1~km of a negative leader from Flash B. This represents only a small section of the full flash that covers an area of the order $25 \times 10$~km$^2$ and is active for 2~s. For the TRID image (\figref{Intf-NLaH}, right) only strong sources ($I>30$~gb) are plotted resulting in 399 sources while the quality cuts on the impulsive image (left) leaves 159 sources. The two imaging methods are subject to different systematic and statistical uncertainties and have a different resolution. One should therefore focus on general structures and not on individual points and as is seen from \figref{Intf-NLaH}, there is no significant difference between the two. The sources in the TRID image show a bit less scatter than those from the impulsive imager but the differences are marginal.

We observe that the received coherent power integrated over 0.3~ms for a section of a  negative leader of Flash A, located at the NE of the array, is about twice as large for the Y-dipoles as it is for the X-Dipoles. For Flash B (at the SE of the core) the intensity ratio is reversed. This is due to the azimuth-angle dependence of the antenna gain. For sources in the NE horizontally polarized radiation is recorded almost exclusively in the Y-dipoles with a gain (in power) that is about a factor two larger than that for vertical polarization recorded exclusively by the X-dipoles in this configuration while for sources in the SE the ratio is reversed. From the observation of the intensity ratio for X- and Y-dipoles one thus can conclude that on average there is no preferred polarization direction for the VHF emission from a negative leader.

\subsection{Detailed view of a negative leader}\seclab{Dynamics}

\begin{figure*}[th]
\centering{\includegraphics[ bb=1.0cm 2.4cm 24.5cm 30.4cm,clip, width=0.49\textwidth]{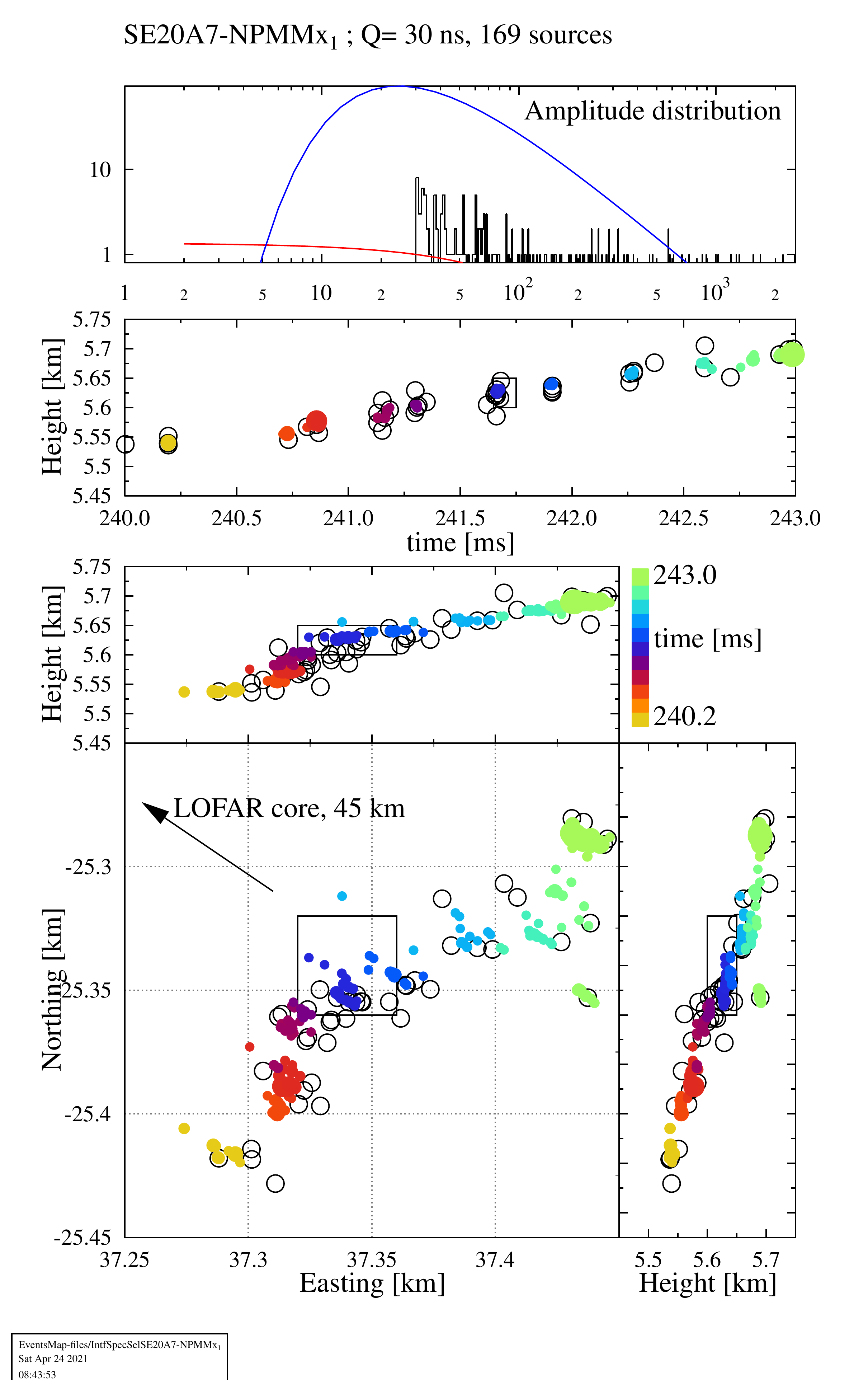} }
\centering{\includegraphics[ bb=1.0cm 2.4cm 24.5cm 30.4cm,clip, width=0.49\textwidth]{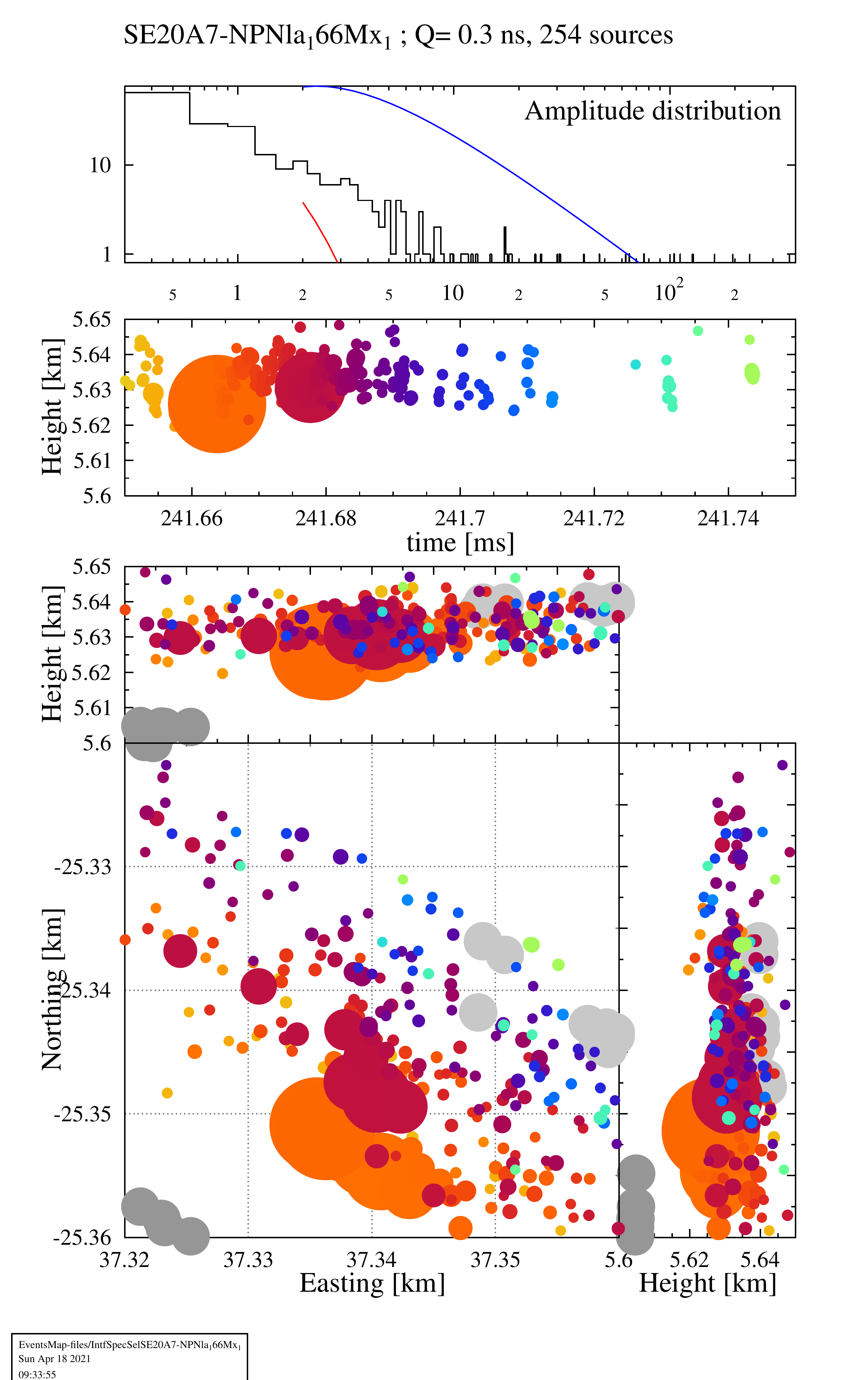} }
	\caption{Left panels show a zoom in on a section of the negative leader shown in \tabref{NegLead}. The sources found by the impulsive imager (44 sources) and in color those from TRID with a $I>30$~gb (99 sources) are indicated by black open circles. The arrow indicates the radial direction and points to the core of LOFAR. The right shows an even further zoom-in on one particular burst as imaged using TRID with an intensity threshold of $I>0.3$~gb showing 153 sources. In light (darker) grey the interferometric more intense sources are shown that fall after (before) the time frame of the plot. The area of the dots is proportional to the intensity for the colored strong sources. \figlab{Intf-NLaZM}}	
\end{figure*}

The strength of TRID-imaging is shown in \figref{Intf-NLaZM} where zoom-ins are made at two different levels on the leader track shown in \figref{Intf-NLaH}. The left side of \figref{Intf-NLaZM} shows the stronger sources ($I>30$~gb) only, while for the zoom-in on the right all sources with an intensity $I>0.3$~gb (i.e.\ with a strength below the noise level of a single antenna) are shown.

One aspect evident from \figref{Intf-NLaZM} is that our imaging accuracy is very directional dependent, as was remarked already in \secref{Vox}. Sources occurring in a narrow time window, i.e. have the same color, show a much larger spread in the ($\hat{R}$) direction (radially outward from the core at (0,0), indicated by the black arrow on the left plan view of \figref{Intf-NLaZM}) than in the transverse direction, about a factor 10 difference.

\figref{Intf-NLaZM} shows clearly that the real strength of TRID imaging lies in its ability to image the finer details of a flash.
The time-colored sources on the left of \figref{Intf-NLaZM} are obtained through TRID imaging while the sources marked by black circles have been found using the impulsive imager, using the same quality and interferometric-intensity cuts as in \figref{Intf-NLaH} for both. The top left panel clearly shows that also this leader has a distinct burst structure that is typical for negative leader propagation and was analyzed extensively in \cite{Hare:2020}.
The right panels of \figref{Intf-NLaZM} show a zoom in on the  burst indicated by the black boxes on the left side. The size of the colored circles reflect source intensity. The limit on the intensity has been decreased to 0.3~gb, well below the noise level in the reference antenna, which is the reason that many more sources show (a total of 254 for the zoomed-in section). As is to be expected, the positioning accuracy of the weaker sources is less and they thus show more scatter, in particular in the radial direction. The grey sources in the right panels mark the positions of intense TRID-imaged sources that fall outside the imaged time span of 241.65 -- 241.75~ms used for the colored sources. The earlier sources are darker grey (at time $t < 241.31$~ms) and the lighter ones occur later than the colored sources (at $t>241.89$~ms). For all grey sources the dot-size is constant.

In the plan view (\figref{Intf-NLaZM}, bottom right) the leader propagates from the lower left corner to the middle right, i.e.\ from the dark grey dots to the light grey ones. The height v.s.\ time panel, the top of the right side of \figref{Intf-NLaZM}, shows that the burst itself has a very distinct structure, starting with some intense sources, the corona flash itself, followed by weaker sources with a decreasing density. From the plan view, the bottom panel, it is seen that they follow the propagation direction of the leader, as expected for streamers ensuing from a corona flash.
Unfortunately the image quality is worse in the radial direction (almost perpendicular to the propagation direction of the negative leader) and it thus cannot be seen if the streamer activity occurs in a narrow forward cone or in a broad fanned-out structure. The propagating ionization front has a width (in the propagation direction) of less than 5~m which is about the locating accuracy we reach for these small intensity sources. The streamers from the flash took 100~$\mu$s to propagate over a distance of about 20~m. The average velocity is about $2\times 10^5$~m/s, but it can be seen that the velocity of the streamers has dropped considerably towards the end of their reach. At this distance the following corona flash starts, as indicated by the light grey dots, some 160~$\mu$s after this flash ceases to show VHF-emission.

The structure shown at the right of  \figref{Intf-NLaZM} is general for all bursts we see on the two negative leader sections we have investigated for Flash B. 
All burst start with some strong sources and many weaker sources where the number decreases in time. After 100~$\mu$s this has decreased to about the background level and the next burst occurs not more than 0.5~ms later at the furthest distance where streamer activity was observed from the previous flash. For the time period between the end of one burst and the start of the next the area is VHF-quiet and we detect only background fluctuations.

The observed structure is very reminiscent of what is observes for High Altitude Negative Leaders (HANL) in \cite{Scholten:2021-HANL} using the impulsive imager. The main difference is that a HANL step length covers a few 100~m while it is only a few 10~m in the present case. In the case of the HANL a filament structure could be distinguished. At the end of the reach of this filaments new coronal bursts were seen to occur, a considerable time after the filament activity had died away.

A possible mechanism is  that in the corona flash some poorly conductive channels (streamers) are formed that stop propagating when a maximal range is reached. Charge, however, continues to move along these streamers to accumulate at the end point. After a certain time, probably depending on conductivity and steamer length, the accumulated charge is sufficient for a break down creating the following corona flash. This would involve a process that may be similar to the stepping seen in laboratory when voltage changes linearly \cite{Kochkin:2014}.
In the current standard picture for the stepping process a space stem grows towards the head of the negative leader where upon contact charge flows towards the outer end of the space stem, heating the space stem to become a leader and initiating a following corona flash. In such a picture one would expect some VHF emission from the space stem when the sudden increase in (negative) charge causes a sudden drop in its potential. In this initial work we see absolutely no evidence for this process, which needs to be investigated further in future work.

The spatial development of corona flashes could not be distinguished as well on the negative leaders for Flash A, although it is seen for some.  This might be related to the fact that the investigated negative leaders for Flash A are at a lower altitude than for Flash B. Combined with the observations in \cite{Scholten:2021-HANL} this would support a strong altitude dependence of the stepping process, as was already suggested in \cite{WuT:2015, Lyu:2016}. We will investigate the occurrence of corona flashes in a future, more extensive study of negative leaders using  TRID imaging.

\subsection{Intensity spectrum}\seclab{PowerLaw}

\begin{figure}[th]
\centering{\includegraphics[ bb=0.2cm 2.0cm 22cm 15cm,clip, width=0.49\textwidth]{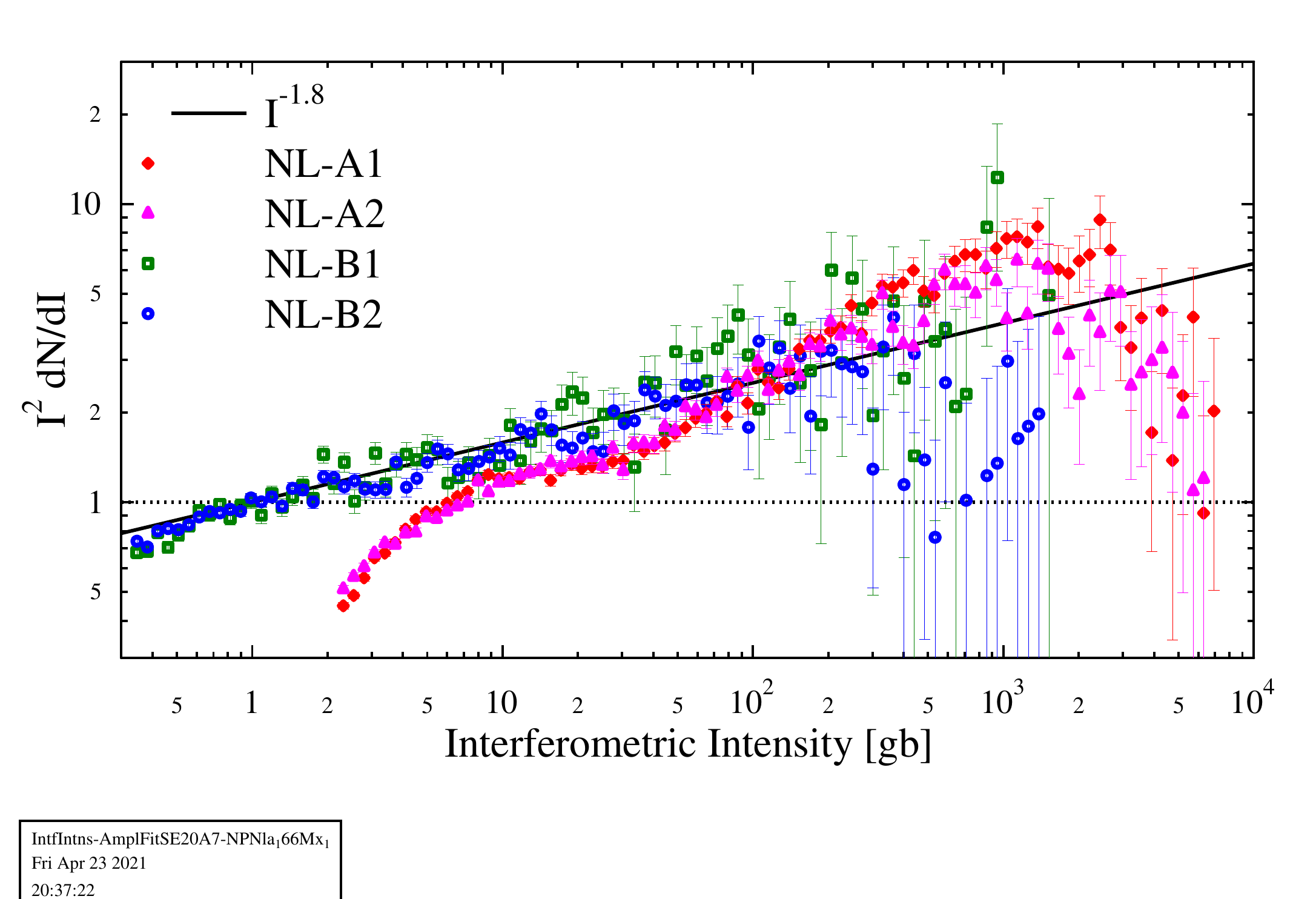} }
	\caption{Distribution of source intensities for the four negative leaders sections given in \tabref{NegLead}. The analysis shown is using the Y-dipoles only. A power law dependence is shown to guide the eye.}	 \figlab{IntfIntS}
\end{figure}

The impulsive imager can reconstruct the source location of a pulse only when it is clearly distinct in the time trace. In practice this means that only for the strongest pulses the source locations can be reconstructed and the reconstruction efficiency drops rapidly with pulse density. This imaging efficiency is difficult to quantify. Since the TRID imager does not rely on pulse finding, the spectrum slicing is independent of the pulse structure, the efficiency for reconstructing the source locations of the received intensity is almost 100\% for sources in the image cube, with the provision  that at most a single source position can be reconstructed within the time frame of a slice.
In practice this means that (almost) all sources are reconstructed when the density of sources in time is less than the inverse of the slicing time. Another provision is that there should  be no strong sources in the vicinity of the image cube since otherwise the images are blurred by the side beams from these strong sources.
This completeness of the spectrum allows for an analysis of the intensity spectrum.

The distribution of the interferometric source intensities (see \secref{Cal}) is given in \figref{IntfIntS} using a slicing time of 100~ns. The distributions are calculated for the four negative leader sections listed in \tabref{NegLead}. The interferometry analysis is performed separately for the X- and Y-dipoles that have different orientations and thus different polarization sensitivities. The extracted distributions are very similar for the two dipoles and only the results for the Y-dipole analysis is given in \figref{IntfIntS}.
The plotted distributions have been normalized (by dividing by a factor $F=N_t \, I_{\rm Th}$ where $N_t$ is the total number of sources in the distribution and $I_{\rm Th}$ is the threshold for including sources) to express the similarity for the different leaders more clearly. Additionally the plotted distributions have been scaled by the square of the intensity in order to emphasize smaller details.
The factor $F$ is chosen such that a perfect inverse square power law dependence will show as a horizontal line at unity.

The intensity distributions for the four negative leader sections are all very similar as is shown in \figref{IntfIntS}. The distributions follow a power law, and to guide the eye an $I^{-1.8}$ dependence is shown by the black line.
At smaller intensities the distributions saturate and even drop off to zero because at most a single source, the strongest, is located per time slice of 0.1~$\mu$s. The weaker sources are therefore `masked' by the stronger ones and the power spectrum is no longer complete. The main differences between the distributions show in the saturation at small intensities and is related to the measured power. The measured power depends on the altitude of the source as well as the distance, see \secref{AntWeight} for a discussion, and differs for X- and Y-dipoles.
At the largest intensities there are signs of a drop-off in the power law, however this may also be due to limited statistics.

\begin{table}[ht]
\centering
\begin{tabular}{||c||c|c|c||}
\hline
Leader & ${\cal N}$ & $\alpha$ & $\gamma$ \\
\hline
NL-A1-Y&   0.88 &  1.76 &    1.49 \\
NL-A1-X &  0.89 &  1.79 &    1.04 \\
\hline
NL-A2-Y &  0.85 &  1.75 &    1.24 \\
NL-A2-X &  0.87 &  1.78 &    1.09 \\
\hline
NL-B1-Y &  1.3 &  1.85 &     0.110 \\
NL-B1-X &  1.0 &  1.81 &     0.135 \\ 
\hline
NL-B2-Y &  1.3 &  1.88 &     0.102 \\
NL-B2-X &  1.5 &  1.90 &    0.220 \\ 
\hline
\end{tabular}
\caption{The values for the normalization ${\cal N}$, the power $\alpha$, and the small-intensity suppression coefficient $\gamma$ in \eqref{PowLaw} as extracted for fits to the intensity distributions for the for negative leader sections listed in \tabref{NegLead}. Separate analyses are made for the X- and Y-dipoles. \tablab{IntAmp}}
\end{table}

To account for the suppression at small intensities we fit the normalized distributions $N(I)$ with a modified power law,
\eqref{PowLaw}.
\begin{equation}
N(I)= {\cal N} \,I^{-\alpha}  \,e^{-\gamma/I}\;, \eqlab{PowLaw}
\end{equation}
where $I$ is expressed in units of [gb]. The last factor, dependent on $\gamma$, suppresses the distribution at small amplitudes in good agreement with the data. The values for the fitted normalization ${\cal N}$, the power $\alpha$, and the small-intensity suppression factor $\gamma$ are given in \tabref{IntAmp} for four negative leader sections given in \tabref{NegLead} and the two different dipoles.

The fitted powers $\alpha$ are almost the same for all eight cases. The main differences are seen in the values for $\gamma$ indicating that for Flash A the Y-dipoles have a larger gain than the X-dipoles while for Flash B the situation is reversed, see the discussion in \secref{LargeScale}. Since Flash A occurs closer to the core, the $\gamma$ coefficients for Flash A are all larger than those for B. In a future work we will investigate the extent of the power law dependence in more detail, in particular if this applies to all parts of the flash and the cut-off in the power law which one would expect at the highest intensities since the total radiated energy should not diverge.

It is interesting to note that power-law dependencies are usually related to disruptive processes, \cite{MengFZ:2019,Stumpf:2012,Virkar:2014}, and in this respect it should have been no surprise to find this in association with a corona flash. On the other hand it is surprising to observe a power-law, which is scale invariant, as one is dealing with a process where many scales enter such as the length scale that is given by the stepping length, the time scale set by the stepping time, or the scale set by the total charge in the leader tip.

A power-law dependence was observed already in \cite{Machado:2021} using the data from the impulsive imager where an unbiased intensity spectrum could be obtained for the 10\% most intense sources.

\section{Summary and outlook}\seclab{summ}

Time Resolved Interferometric 3D-imaging gives an unprecedented insight in the finer dynamics of lightning propagation. Using hundreds of antennas with baselines of up to 100~km allows to image the more weakly emitting structures with (a better than) 10~m scale precision. The TRID imaging procedure should be considered complementary to the impulsive imager presented in \cite{Scholten:2021-init} that builds an image by locating the sources of single, reasonably well isolated, pulses in the spectrum. The impulsive imager can efficiently image the complete flash while the TRID-imager allows to image much fainter aspects of the flash as well as a somewhat improved locating accuracy, however  at the expense of being much more CPU-intensive.

As a first, exploratory, application of TRID imaging we have applied it to observe the finer details of negative leader bursts and imaged the temporal and spatial development of a corona flash. We show that it expands up to the  distance where the flash of the following step starts. In addition we show that the intensities of VHF pulses emitted by negative leaders follow an almost perfect inverse-square power law extending to the largest intensities. Each of these subjects requires a much more in-depth analysis which we plan for forthcoming publications.

In spite of the unprecedented possibilities of the present method to study the fine dynamics of lightning, there are still several aspects where the method should be improved. A minor improvement would be to optimize the antenna-weighting scheme discussed in \secref{AntWeight} or implement a more sophisticated construction of the position of the maximum (or maxima), see \secref{Max}. An important improvement is foreseen through the implementation of polarization, angle, and frequency dependence of the antenna gain when performing the coherent sum over all antennas. Apart from lifting the requirement that the flash is seen at small elevation angles and in a not too large azimuth angle range, see \secref{Cal}, this has the potential of improving the imaging quality, giving access to the direction of the electrical current moment in the VHF-source, and allowing the implementation of a true de-convolution of the beam structure giving the possibility to locate multiple sources per window.

\section{acknowledgements}
The project has also received funding from the European Research Council (ERC) under the European Union's Horizon 2020 research and innovation programme [grant number  640130]; 
We furthermore acknowledge financial support from FOM [FOM-project 12PR304]; 
BMH is supported by the NWO [grant number VI.VENI.192.071];
AN is supported by the DFG [grant number NE 2031/2-1];
\\
This paper is based on data obtained with the International LOFAR Telescope (ILT). LOFAR~\cite{Haarlem:2013} is the Low Frequency Array designed and constructed by ASTRON. It has observing, data processing, and data storage facilities in several countries, that are owned by various parties (each with their own funding sources), and that are collectively operated by the ILT foundation under a joint scientific policy. The ILT resources have benefitted from the following recent major funding sources: CNRS-INSU, Observatoire de Paris and Universit\'{e} d'Orl\'{e}ans, France; BMBF, MIWF-NRW, MPG, Germany; Science Foundation Ireland (SFI), Department of Business, Enterprise and Innovation (DBEI), Ireland; NWO, The Netherlands; The Science and Technology Facilities Council, UK.

\section{Data statement}\seclab{data}

The data are available from the LOFAR Long Term Archive (for access see \citep{LOFAR-LTA}).
To download this data, please create an account and follow the instructions for ``Staging Transient Buffer Board data'' at \citep{LOFAR-LTA}. In particular, the utility ``wget'' should be used as follows:
\\ {\footnotesize \verb!wget https://lofar-download.grid.surfsara.nl/lofigrid!}
\\\hspace*{7em} {\footnotesize \verb!/SRMFifoGet.py?surl="location"!}
\\where ``location'' should be specified as:
\\ {\small \verb!srm://srm.grid.sara.nl/pnfs/grid.sara.nl/data/lofar!}
\\\hspace*{7em} {\small \verb!/ops/TBB/lightning/!} \\ followed by
\\ {\small \verb!L668464_D20180921T194259.023Z_"stat"_R000_tbb.h5!} (for Flash A)
\\ {\small \verb!L792864_D20200814T143241.768Z_"stat"_R000_tbb.h5!} (for Flash B)
\\ {\small \verb!L792864_D20200814T143241.768Z_"stat"_R000_tbb.h5!} (for Flash C)
and where
``stat'' should be replaced by the name of the station, CS001, CS002, CS003, CS004, CS005, CS006, CS007, CS011, CS013, CS017, CS021, CS024, CS026, CS028, CS030, CS031, CS032, CS101, CS103, RS106, CS201, RS205, RS208, RS210, CS301, CS302, RS305, RS306, RS307, RS310, CS401, RS406, RS407, RS409, CS501, RS503, RS508, or RS509.
The source code used for TRID imaging can be found at \citep{Scholten_LofarImaging:2020, Scholten-v19:2021}.
All figures in this work have been made using the Graphics Layout Engine (GLE) \cite{GLE} plotting package.

%
\bibliography{}

\end{document}